\documentclass[aps,twocolumn,showpacs,superscriptaddress]{revtex4}
\usepackage{graphicx,dcolumn}% Include figure files & Align table columns on decimal point
\usepackage{bm,amsmath,amssymb}
\usepackage{times}
\usepackage[colorlinks,citecolor=blue,linkcolor=blue]{hyperref}

\begin{document}
\title{Dynamic control of quantum geometric heat flux in a nonequilibrium spin-boson model}
\author{Tian Chen}
 \affiliation{State Key Laboratory of Low
Dimensional Quantum Physics, Department of Physics, Tsinghua University, Beijing 100084,
People's Republic of China}
\author{Xiang-Bin Wang}
 \email{xbwang@mail.tsinghua.edu.cn}
 \affiliation{State Key Laboratory of Low
Dimensional Quantum Physics, Department of Physics, Tsinghua University, Beijing 100084,
People's Republic of China}
\affiliation{ Jinan Institute of Quantum Technology, Shandong
Academy of Information and Communication Technology, Jinan 250101,
People's Republic of China}
\author{Jie Ren}
\email{renjie@lanl.gov}
\affiliation{Theoretical Division, Los Alamos National Laboratory, Los Alamos, New Mexico 87545, USA}

\begin{abstract}
We study the quantum geometric heat flux in the nonequilibrium spin-boson model. By adopting the noninteracting-blip approximation that is able to accommodate the strong system-bath coupling, we show that there exists a nonzero geometric heat flux only when the two-level system is nondegenerate.
Moreover, the pumping, no pumping, and dynamic control of geometric heat flux are discussed in detail, compared to the results with Redfield weak-coupling approximation. In particular, the geometric energy transfer induced by modulation of two system-bath couplings is identified, which is exclusive to quantum transport in the strong system-bath coupling regime.
%\\Keywords: adiabatic modulation; geometric heat current; NIBA method
\end{abstract}

% 44.90.+c	Other topics in heat transfer (restricted to new topics in section 44)

\maketitle

\section{Introduction} 
Smart energy control in low-dimensional nanoscale systems is of both theoretical and practical importance, rendering the emergence of Phononics  \cite{N. B. Li}, where various functional thermal devices are designed for managing thermal energy and information at nanoscale. As is well known, according to the second law of thermodynamics, energy cannot transfer from a cold reservoir to a hot side spontaneously without external modulation. In order to obtain a more flexible control of thermal energy at the nanoscale, there is a great demand for designing intriguing phononic thermal devices, which can utilize temporal modulations to achieve dynamic control, such as in heat pumps, motors and engines.

So far, many proposals have been provided to dynamically control the energy flow between a cold reservoir and a hot one \cite{D. Segal, A. Dhar, LiEPL, W. Zhang, S. Zhang, A. Dhar2, Kovhan, S. Narayana}. As a result of these investigations \cite{S. Zhang, A. Dhar}, there is no way to pump the energy from the cold side to the hot side in classical-oscillator systems by force driving, though heat pump can be implemented in classical spin chains \cite{A. Dhar}. An interesting design to realize the dynamic control of heat transfer utilizes the adiabatic geometric phase effect \cite{J. Ren, S. Liu}. Similar to the geometric phase in a closed quantum system under adiabatic drivings, when an open system with reservoirs is subjected to time-dependent modulations, the energy transfer will also gain such a geometric-phase-induced additional energy flux  \cite{J. Ren, S. Liu}.
%In particular, Ref. \cite{J. Ren} proposed to pump the energy between two reservoirs with equal temporal-averaged-temperature by utilizing the geometric phase induced heat transfer. Very recently, similar geometric heat flux in classical interacting systems is also studied \cite{S. Liu}.

The previous study of geometric heat flux in a spin-boson model adopted the Redfield weak-coupling scheme \cite{J. Ren}.
There are many approximation methods in studying the heat transfer in the spin-boson model \cite{A. Nitzan, L. A. Wu, K. A. Velizhanin, Ruokola, L. Nicolin, L. Nicolin2}. Among them, the Redfield weak-coupling approximation is often used since this method is effective \cite{secular}.  The physical picture described by the Redfield weak-coupling scheme has two inherent assumptions. One is the {\em resonant tunneling} between the system and the reservoirs, and only the resonant frequency in the reservoir contributes to the dynamics. The other is that each individual reservoir interacts with the system {\em separately}, or say, {\em additively}.

However, recent studies reveal that there are also limitations to the Redfield-weak coupling scheme \cite{K. A. Velizhanin, ChenWang, L. Nicolin, L. Nicolin2, F. Nesi}. Besides the scheme, other methods such as the multilayer multiconfiguration Hartree \cite{K. A. Velizhanin} and the noninteracting-blip approximation (NIBA) \cite{Legget, H. Dekker, Weissbook} are also applied to study the heat flow in a spin-boson model under temperature bias \cite{L. Nicolin, L. Nicolin2}. % Interestingly, these methods can produce  results different from that of the Redfield approximation method for the same Hamiltonian. 
Particularly, as shown in Refs. \cite{K. A. Velizhanin, L. Nicolin2}, the heat current is not linearly dependent on the system-reservoir coupling strength as given by the Redfield weak-coupling scheme. There exists a maximal heat current at the intermediate system-reservoir coupling regime, and then the heat current decreases in the strong system-reservoir coupling regime. %So far, different physical results of Redfield scheme and NIBA are not found in the weak system-bath interaction regime.

In this paper,  we adopt the NIBA scheme to study the geometric phase-induced energy transfer in a spin-boson model. Distinct from the Redfield scheme, NIBA is well known as a scheme accommodating the strong system-reservoir coupling. It describes another different physical scenario of transport \cite{L. Nicolin, L. Nicolin2}: {\it nonresonant} tunneling between the system and the reservoirs, and the {\it collective nonadditive} interacting between reservoirs. We shall first review the analytical expressions of the investigated spin-boson model in NIBA and the geometric phase induced energy transfer through the generating function approach. We then investigate in detail the behaviors of geometric heat flux under various modulation protocols. 

Our contributions are twofold.
First, through calculating the geometric phase induced heat flux in unbiased (degenerate) case, we find zero geometric heat flux for NIBA method here, which is different from the finite geometric heat flux for Redfield scheme \cite{J. Ren}. Although, Ref. \cite{L. Nicolin2} showed that even in the weak coupling limit, NIBA method agrees very well with Redfield calculation of heat flux, we can see that they still describe distinct ((non)-resonant and (non)-additive) physical pictures for nonequilibrium energy transfer in the spin-boson model.
Our results indicate that the geometric heat flux is a sensitive indicator for different physical pictures of transfer dynamics, which renders geometric heat flux could become an effective tool to compare the difference or even judge the correctness of various approximation schemes.

Second, geometric heat flux itself is a very important physical problem. A thorough investigation into the problem with different control protocols is meaningful.  
We find that geometric heat flux under two-temperature modulations is linearly increasing with the system-bath coupling strength for the biased  (nondegenerate) case. For the modulation of one temperature and the two-level energy gap, we also observe a similar increase of the geometric heat flux with the system-bath coupling. This is useful because the dynamic heat flux always decays in the strong system-bath coupling limit, so the geometric heat flux will dominate the energy transfer in that strong-coupling limit and will have detectable consequences.  Moreover, we find the direction of geometric heat flux can be reversed by adjusting the system's parameters, in addition to reversing the modulation protocol. Finally, two-coupling-modulation-induced geometric heat flux is identified. So far, this nontrivial observation is exclusive for quantum transport in the strong system-bath coupling regime.

%\begin{figure}%[t]
%%\vspace{-3cm}
%%\hspace{-1cm}
%\scalebox{0.45}[0.45]{\includegraphics{newfig1.pdf}}
%\vspace{-.3cm}
%\caption{(color online). Schematic illustrations of the nonequilibrium spin-boson model for energy transfer.}
%\label{fig:scheme}
%\end{figure}

\section{Model} 
The nonequilibrium spin-boson (NESB) model, %[see the left panel of Fig. \ref{fig:scheme}], 
consisting of a two-level system in contact with two bosonic reservoirs with temperatures $T_{\nu}(\nu=L,R)$, is described by the Hamiltonian:
\begin{equation}
H=\frac{\varepsilon_{0}}{2}\sigma_{z}+\frac{\Delta}{2}\sigma_{x}+\sigma_{z}\sum_{\nu,j}\lambda_{j,\nu}(b_{j,\nu}^{\dag}+b_{j,\nu})+\sum_{\nu,j}\omega_{j,\nu}b_{j,\nu}^{\dag}b_{j,\nu},
\label{eq:H1}
\end{equation}
where $\varepsilon_{0}$ is the energy gap of the two levels; $\Delta$ denotes the tunneling between them; $\sigma_z\equiv|1\rangle\langle1|-|0\rangle\langle0|$ and $\sigma_x\equiv|0\rangle\langle1|+|1\rangle\langle0|$ are the Pauli matrices expressed in the two level basis; $b^\dag_{j,\nu} (b_{j,\nu})$ denotes the creation (annihilation) operator of the $j$th harmonic mode in the $\nu$ bosonic bath, with $\lambda_{j,\nu}$ the system-bath coupling strength. 
%We can also rotate the Hamiltonian (\ref{eq:H1}) firstly around the $y$ axis by an angle $\pi/2$, then around the new $x$ axis by an angle $\pi$. In this way, $\sigma_x\rightarrow\sigma_z, \sigma_z\rightarrow\sigma_x$, and the alternative equivalent Hamiltonian reads, [see the right panel of Fig. \ref{fig:scheme}]:
%\begin{equation}
%H=\frac{\varepsilon_{0}}{2}\sigma_{x}+\frac{\Delta}{2}\sigma_{z}+\sigma_{x}\sum_{\nu,j}\lambda_{j,\nu}(b_{j,\nu}^{\dag}+b_{j,\nu})+\sum_{\nu,j}\omega_{j,\nu}b_{j,\nu}^{\dag}b_{j,\nu}.
%\label{eq:H2}
%\end{equation}
%Throughout the paper, we will work in the basis of Hamiltonian (\ref{eq:H1}) in the NIBA scheme.
Before proceeding to the energy transport calculations, it is useful to transform the NESB Hamiltonian (\ref{eq:H1}) by using the canonical transformation \cite{Mahanbook, LF} (also called Lang-Firsov or polaron transformation): $H_{T}=U^{\dag}HU$, $U=\exp[{i\sigma_{z}\Omega/2}]$, $\Omega=\sum_{\nu}\Omega_{\nu}=2i\sum_{\nu,j}\frac{\lambda_{j,\nu}}{\omega_{j,\nu}}(b_{j,\nu}^{\dag}-b_{j,\nu})$. After neglecting an irrelevant constant $-\sum_{j,\nu}{\lambda^2_{j,\nu}}/{\omega_{j}}$, the transformed Hamiltonian is expressed as:
\begin{equation}
H_{T}=\frac{\varepsilon_{0}}{2}\sigma_{z}+\frac{\Delta}{2}(\sigma_{+}e^{-i\Omega}+\sigma_{-}e^{i\Omega})+\sum_{j,\nu}\omega_{j,\nu}b^{\dag}_{j,\nu}b_{j,\nu},
\end{equation}
with $\sigma_{+}\equiv(\sigma_x+i\sigma_y)/2=|1\rangle\langle0|$ and $\sigma_{-}\equiv(\sigma_x-i\sigma_y)/2=|0\rangle\langle1|$. $H_T$ clearly shows that the energy transfer is accomplished by the excitation from the lower level to the upper one with absorbing energy from the baths, and the relaxation from the upper level to the lower one with releasing energy to the baths. Thereafter, as a result of the NIBA method \cite{F. Nesi, Legget, H. Dekker, Weissbook} and with the Markov assumption, the population dynamics of the two-level system becomes \cite{L. Nicolin2}:
\begin{equation}
\frac{d}{dt}\left(
\begin{array}{c}
p_0(t)\\
p_1(t)
\end{array}
\right)=
-\left(
\begin{array}{cc}
K(-\varepsilon_{0})  & -K(\varepsilon_{0})\\
-K(-\varepsilon_{0}) & K(\varepsilon_{0})
\end{array}
\right)
\left(
\begin{array}{c}
p_0(t)\\
p_1(t)
\end{array}
\right),
\label{eq:ME1}
\end{equation}
where, $p_{0/1}(t)\equiv (1\mp \langle\sigma_z(t)\rangle)/2$ denotes the population at the lower (upper) level. The transition rates stand for the cooperative process between the system and two reservoirs:
\begin{equation}
\begin{split}
K(\pm\varepsilon_{0})&\equiv\frac{({\Delta}/{2})^2}{2\pi}\int_{-\infty}^{\infty}C^{\pm}(\omega)d\omega, \\ 
C^{\pm}(\omega)&\equiv C_{L}(\pm\varepsilon_{0}\mp\omega)C_{R}(\pm\omega), 
\end{split}
\label{eq:rates}
\end{equation}
with $C_{\nu}(\omega)\equiv\int_{-\infty}^{\infty}e^{i\omega t-Q_{\nu}(t)}dt$ denoting the probability density of the bath $\nu$ to absorb the energy $\omega$ (equivalently, to release the energy $-\omega$).
Employing the Gaussian statistics of the ``momentum'' operator $\Omega(t)$, we have $Q_{\nu}(t)\equiv\langle[\Omega_{\nu}(0)-\Omega_{\nu}(t)]\Omega_{\nu}(0)\rangle=\int_{0}^{\infty}\frac{J_{\nu}(\omega)}{\pi \omega^{2}}\big[\coth(\frac{\omega}{2T_{\nu}})\left(1-\cos(\omega t)\right)+i\sin(\omega t)\big]d\omega$ \cite{Weissbook}, with $J_{\nu}(\omega)=4\pi\sum_{j}\lambda^2_{j,\nu}\delta(\omega-\omega_{j,\nu})$ being the spectral density of the bosonic bath $\nu$.  In contrast to the Redfield-weak coupling scheme \cite{J. Ren, A. Nitzan}, the rate expressions (\ref{eq:rates}) distinctly exhibit the non-resonant energy tunneling processes, conjoining the two baths nonadditively: $K(\varepsilon_0)$ means when the central system loses energy $\varepsilon_0$ by relaxing from the upper level to the lower one, the $R$ bath will absorb $\omega$ and the $L$ bath gains the rest if $\varepsilon_0>\omega$ or even supplements the shortage if $\omega>\varepsilon_0$; $K(-\varepsilon_0)$ depicts a similar energy transfer process for the central system exciting from the lower level to the upper one.

\section{Generating function and geometric heat flux} 
Following the {\it Full Counting Statistics} \cite{M. Esposito, M. Campisi}, we next construct the cumulant generating function (CGF) of the NESB model to count the phonon energy transfer through the right system-bath coupling \cite{ChenWang, J. Ren, S. Liu, Bagrets, Gopich, N. A. Sinitsyn}. Denote ${\rho}_{t}(n,\omega)$ as the joint probability that a total energy of $\omega$ has been transferred to the right bath during time interval $[0, t]$, with the two-level system populates at state $|n\rangle$ ($n=0,1$) at time $t$, we then introduce the characteristic function of  that joint probability $|z(\chi,t)\rangle\equiv(\int_{-\infty}^{\infty}{\rho}_{t}(0,\omega)e^{i\omega\chi}d\omega, \int_{-\infty}^{\infty}{\rho}_{t}(1,\omega)e^{i\omega\chi}d\omega)^T$ with the energy counting field $\chi$. Following \cite{L. Nicolin2}, this characteristic function satisfies the following dynamics:
\begin{equation}
\begin{split}
\frac{d}{dt}|z(\chi,t)\rangle&=-\hat{\mathcal{H}}(\chi)|z(\chi,t)\rangle, \label{eq:ME2} \\
\text{with} \quad \hat{\mathcal{H}}(\chi)&=\left(
\begin{array}{cc}
K(-\varepsilon_{0})  & -K^{+}(\chi) \\
-K^{-}(\chi) & K(\varepsilon_{0})
\end{array}
\right),
\end{split}
\end{equation}
where $K^{\pm}(\chi)\equiv\frac{({\Delta}/{2})^2}{2\pi}\int_{-\infty}^{\infty}C^{\pm}(\omega)e^{\pm i\omega\chi}d\omega$. When the counting field $\chi=0$, $K^{\pm}(0)=K(\pm\epsilon_0)$ and in turn Eq. (\ref{eq:ME2}) reduces to Eq. (\ref{eq:ME1}). Thus, the characteristic function of the heat transfer is $Z(\chi, t)=\int_{-\infty}^{\infty}\big({\rho}_{t}(0,\omega)+{\rho}_{t}(1,\omega)\big)
e^{i\omega\chi}d\omega$ and the CGF is $\mathcal{G}(\chi)\equiv\lim_{t\rightarrow\infty}\frac{1}{t}\ln[Z(\chi, t)]$, which generates the $n$-order cumulant of heat transfer fluctuations through $\lim_{t\rightarrow\infty}\langle\langle Q^n\rangle\rangle/t=\partial^n_{i\chi}\mathcal{G}(\chi)|_{\chi=0}$. The mean value of the heat flux is just the first order: $J=\partial_{i\chi}\mathcal{G}(\chi)|_{\chi=0}$

Behaviors in the long-time limit are of our central interest. They are governed by the ground state of the operator $\hat{\mathcal{H}}(\chi)$, of which the eigenvalue $E_0(\chi)$ possesses the the smallest real part. For time-independent $\hat{\mathcal{H}}(\chi)$, $\lim_{t\rightarrow\infty}Z(\chi, t)\sim e^{-E_0(\chi)t}$ and in turn $\lim_{t\rightarrow\infty}\langle\langle Q^n\rangle\rangle/t=-\partial^n_{i\chi}E_0(\chi)|_{\chi=0}$. However, for time-dependent $\hat{\mathcal{H}}(\chi, t)$, where the system parameters $\Delta(t), \varepsilon_0(t)$, the bath temperature $T_{\nu}(t)$ or the system-bath coupling $\lambda_{j,\nu}(t)$ could be subject to periodic modulations, the adiabatic geometric phase effect has been unraveled to play an important role in the dynamic control of energy transfer \cite{J. Ren, S. Liu}. In this case, although at every instant the dynamics (\ref{eq:ME2}) is preserved, there exist two contributions in the CGF: $\lim_{t\rightarrow\infty}Z(\chi, t)\sim e^{t\mathcal{G}}=e^{t(\mathcal{G}_{dyn}+\mathcal{G}_{geom})}$. One is the dynamic part $\mathcal{G}_{dyn}$, and the other is the geometric part $\mathcal{G}_{geom}$ \cite{J. Ren, S. Liu, N. A. Sinitsyn}:
\begin{equation}
\begin{split}
\mathcal{G}_{dyn}&=-\frac{1}{\mathcal{T}_{p}}\int_{0}^{\mathcal{T}_{p}}dtE_{0}(\chi,t), \\
\mathcal{G}_{geom}&=-\frac{1}{\mathcal{T}_{p}}\int_{0}^{\mathcal{T}_{p}}dt\langle\phi_{0}|\partial_{t}|\psi_{0}\rangle, 
\end{split}
\end{equation}
with $\mathcal T_p$ the modulating period \cite{Tp} and $|\psi_{0}\rangle (\langle\phi_{0}|)$ the bi-orthonormal right (left) eigenvector corresponding to the ground state of $\hat{\mathcal{H}}(\chi, t)$ \cite{braket}.
In the case of two parameters being modulated, say $u_{1}(t), u_{2}(t)$, the calculation of $\mathcal{G}_{geom}$ can be done using Stokes theorem \cite{A. Bohm},
\begin{equation}
\mathcal{G}_{geom}=-\frac{1}{\mathcal{T}_{p}}\iint_{u_1 u_2}d u_{1}d u_{2}\mathcal{B}_{u_{1}u_{2}}
\end{equation}
and the Berry curvature \cite{A. Bohm, M. V. Berry} is:
\begin{equation}
\begin{split}
\mathcal{B}_{u_{1}u_{2}}&=\langle\partial_{u_{1}}\phi_{0}|\partial_{u_{2}}\psi_{0}\rangle-\langle\partial_{u_{2}}\phi_{0}|\partial_{u_{1}}\psi_{0}\rangle \\
&=\frac{\langle\phi_{0}|\partial_{u_{1}}\hat{\mathcal{H}}|\psi_{1}\rangle\langle\phi_{1}|\partial_{u_{2}}\hat{\mathcal{H}}|\psi_{0}\rangle-(u_{1}\leftrightarrow u_{2})}{(E_{0}-E_{1})^{2}}. 
\end{split}
\label{eq:Berrycurvature}
\end{equation}
with $E_1$ the eigenvalue of the excited state and $|\psi_{1}\rangle (\langle\phi_{1}|)$ the corresponding bi-orthonormal right (left) eigenvector.
According to these formulas, we have the dynamic heat flux $J_{dyn}$ and the geometric heat flux $J_{geom}$,
\begin{eqnarray}
J_{dyn}&=&%\frac{d G_{dyn}(\chi)}{d (i\chi)}|_{\chi=0},\\
-\frac{1}{\mathcal{T}_{p}}\int_{0}^{\mathcal{T}_{p}}dt\left.\frac{\partial E_{0}(\chi,t)}{\partial(i\chi)}\right|_{\chi=0}, \label{eq:Dflux}\\
J_{geom}&=&%\frac{d G_{geom}(\chi)}{d (i\chi)}|_{\chi=0}=
-\frac{1}{\mathcal{T}_{p}}\iint_{u_1u_2}du_1du_2\left.\frac{\partial\mathcal{B}_{u_1u_2}}{\partial(i\chi)}\right|_{\chi=0}.
\label{eq:Berryflux}
\end{eqnarray}

\section{Results and discussions} 
With equations above, we are ready to study the consequences of dynamic control of the NESB model beyond the weak-coupling limit, \emph{i.e.}, without Redfield-weak coupling approximation.

\textbf{i)} {\it Unbiased case, $\varepsilon_{0}=0$.} In this degenerate case, we have $K(-\varepsilon_{0})=K(\varepsilon_{0}), K^{+}(\chi)= K^{-}(\chi)$, and the matrix,
\begin{equation}
\begin{split}
\hat{\mathcal{H}}(\chi, t)&=\left(
\begin{array}{cc}
K(\varepsilon_{0})  & -K^{+}(\chi)\\
-K^{+}(\chi) & K(\varepsilon_{0})
\end{array}
\right).
\end{split}
\end{equation}
Because of the high symmetry of $\hat{\mathcal{H}}(\chi, t)$ at the degenerate case, the corresponding eigenvalues and eigenvectors turn into a simple form,
\begin{equation*}
\begin{split}
&\quad E_{0/1}=K(\varepsilon_{0})\mp K^{+}(\chi), \\
|\psi_{0/1}\rangle=&\frac{1}{\sqrt2}\left(\begin{array}{cc}
\pm 1, & 1\\
\end{array}
\right)^T, \quad  \langle\phi_{0/1}|=\frac{1}{\sqrt2}\left(\begin{array}{cc}
\pm 1, & 1\\
\end{array}
\right).
\end{split}
\end{equation*}
Note now the eigenvectors are constants which indicate that the eigenvectors do not
evolve as the parameter modulations. Thus, from Eq. (\ref{eq:Berrycurvature}), we have that for whatever two-parameter modulations, the Berry curvature $\mathcal{B}_{u_1u_2}\equiv 0$. In other words, there is no geometric phase effect and geometric heat flux is absent when the two-level energy gap is zero. In fact, when $\varepsilon_0=0$, we have $\langle\sigma_z\rangle=0$ and $p_{0/1}(t)\equiv (1\mp \langle\sigma_z(t)\rangle)/2=1/2$. Therefore, nomatter how you drive the system, the two populations always keep constant so that there is no geometric contribution of the transport.

%This result is independent of the system-bath coupling strength. 

\begin{figure}%[t]
%\vspace{-3cm}
%\hspace{-1cm}
\scalebox{0.4}[0.35]{\includegraphics{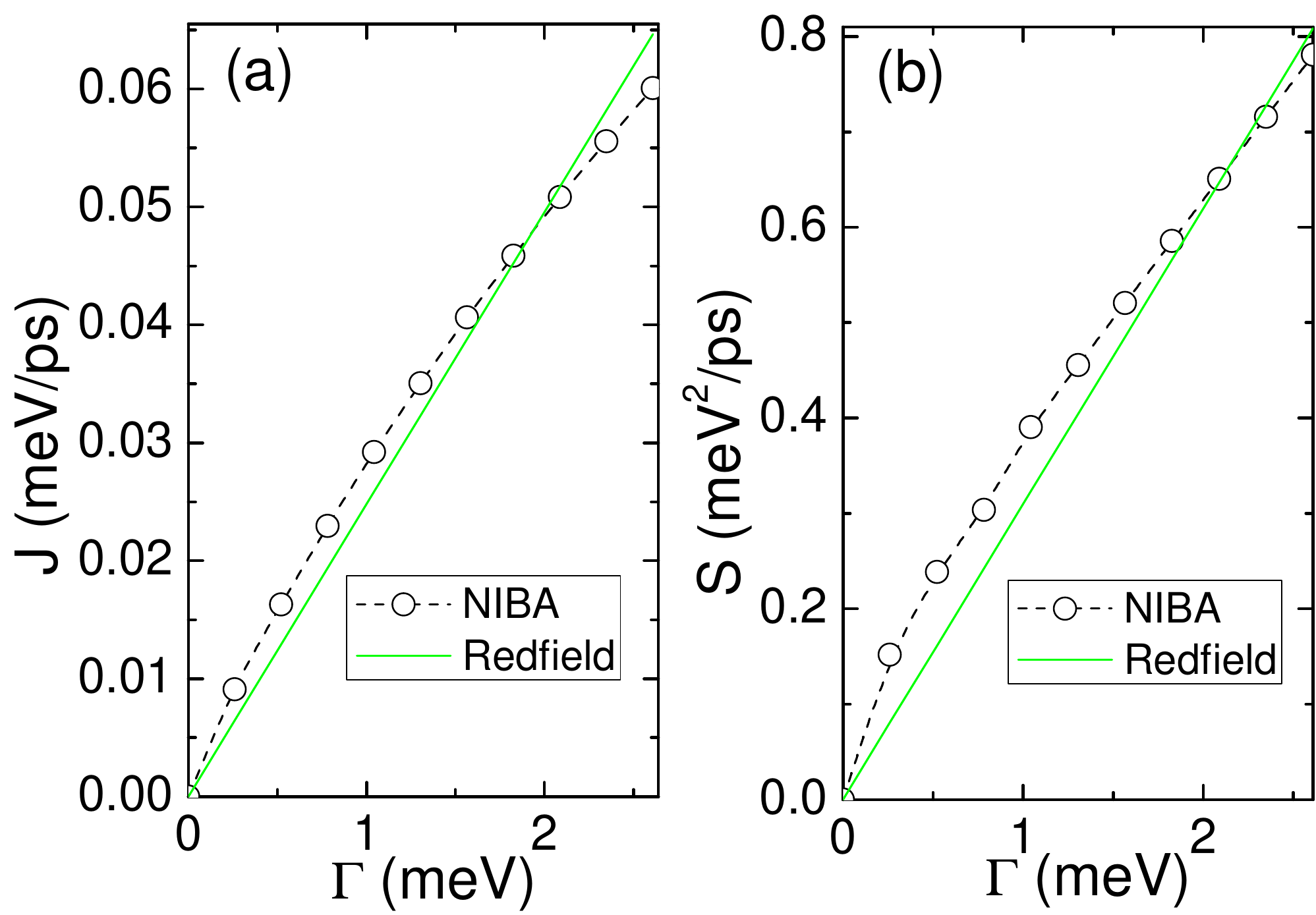}}
\vspace{-.3cm}
\caption{(Color online) Comparison of heat flux and its high-order fluctuations in NIBA and Redfield methods. (a) The mean value of heat flux: $J=-\partial_{i\chi}E_0|_{\chi=0}$. (2) The shot-noise of heat flux: $S=-\partial^2_{i\chi}E_0|_{\chi=0}$. $T_L=150$ K (5), $T_R=90$ K (3), $\Delta=5.22$ meV (2). The cutoff of the Ohmic spectral function is set as $26.1$ meV (10). The numbers in the parentheses are corresponding dimensionless parameters used in Ref. \cite{L. Nicolin2}. }
\label{fig:comp}
\end{figure}

Note that, in contrast to the absence of geometric heat flux here, Ref. \cite{J. Ren} treats the same physical system with $\varepsilon_0=0$ (with notation $\Delta\rightarrow\omega_0$, see also Ref. \cite{L. Nicolin2}), but in the Redfield-weak coupling scheme and finds nonzero geometric heat flux there.  
Despite the difference, these two results are not in conflict with each other. As we have emphasized in the beginning, the NIBA method is not equivalent to the Redfield one. They describe distinct physical pictures of the transport dynamics. Redfield scheme describes the physical picture in the weak coupling limit, which has two inherent assumptions:
1) resonant tunneling, and 2) additive reservoir effect. However, NIBA usually works in the strong coupling regime and describes another different physical picture of transport: 1) off-resonant tunneling, and has 2) non-additive reservoir effect. These differences between the NIBA and Redfield methods give rise to the different results of geometric heat flux. It is just like the two faces of the same coin. The two different faces are not in conflict with each other. They are just two different manifestations of the same system under different scenarios and conditions. 

Rigorously speaking, the NIBA method was originally derived for a spin coupled to a single equilibrium bosonic reservoir. Consequently its theoretical justification for weak-coupling with zero bias \cite{Weissbook} is merely applied for the case of a single bath, where there is no difference between additive and non-additive reservoir effect. Although Ref. \cite{L. Nicolin2} showed that for the transport problem with two baths, even in the weak coupling limit, the NIBA and Redfield methods give heat flux data with similar values, we can see that they still have different curve behaviors [reproduced in Fig. \ref{fig:comp}(a)]. This in fact reflects the two distinct physical pictures described by these two methods \cite{note}. A larger deviation can be found in the shot-noise comparison in Fig. \ref{fig:comp}(b), because the high-order heat flux fluctuations contain more information about the underlying dynamics.

Different from the heat flux and high-order fluctuations, the geometric heat flux contains not only  the information about the ground-state eigenvalues but also more comprehensive information from the eigenfunctions [see Eqs. (\ref{eq:Berrycurvature}), (\ref{eq:Dflux}), (\ref{eq:Berryflux})]. In other words, the geometric heat flux is a more sensitive indicator of the underlying transfer dynamics for different physical pictures. Therefore, we expect that the geometric heat flux could be an effective tool for comparing the differences between or even judging the correctness of various approximation schemes, under different scenarios and conditions.

\textbf{ii)} {\it Biased case, $\varepsilon_{0}\neq0$.} An applied electromagnetic field could control this Zeeman splitting. In this nondegenerate case,  the geometric heat flux  does exist . In the following, to simplify the calculation we assume the Marcus limit \cite{R. A. Marcus, Weissbook} that works at high temperature $T_{\nu}>\varepsilon_{0}$ and/or the strong system-bath coupling regime. The Marcus limit could be approached by a short time expansion of $Q_{\nu}(t)$ such that $Q_{\nu}(t)=\Gamma_{\nu}T_{\nu}t^2+i\Gamma_{\nu}t$ with the renormalized system-bath coupling $\Gamma_{\nu}=\int\frac{J_{\nu}(\omega)}{\pi\omega}d\omega=\sum_j 4\lambda^2_{j,\nu}/\omega_{j,\nu}$. In this way, we have the transition rates \cite{Weissbook, L. Nicolin2}:
\begin{equation*}
\begin{split}
C_{\nu}(\omega)&=\sqrt{\frac{\pi}{\Gamma_{\nu}T_{\nu}}}\exp{\left[-\frac{(\omega-\Gamma_{\nu})^2}{4\Gamma_{\nu}T_{\nu}}\right]}, \\
K(\pm\varepsilon_0)&=\frac{\Delta^2}{4}\sqrt{\frac{\pi}{\Gamma_{L}T_{L}+\Gamma_{R}T_{R}}}\exp{\left[-\frac{(\varepsilon_0\mp\Gamma_{L}\mp\Gamma_{R})^2}{4(\Gamma_{L}T_{L}+\Gamma_{R}T_{R})}\right]},
\end{split}
\end{equation*}
where $\Delta, \varepsilon_0, T_{\nu}, \Gamma_{\nu}$ could be subject to the dynamic control. By substituting these rates into the dynamics. (\ref{eq:ME2}), we are able to investigate the Berry curvature and the geometric heat flux $J_{geom}$ through Eqs. (\ref{eq:Berrycurvature}) and (\ref{eq:Berryflux}).

\begin{figure}%[t]
\vspace{-0.2cm}
%\hspace{-1cm}
\scalebox{0.35}[0.35]{\includegraphics{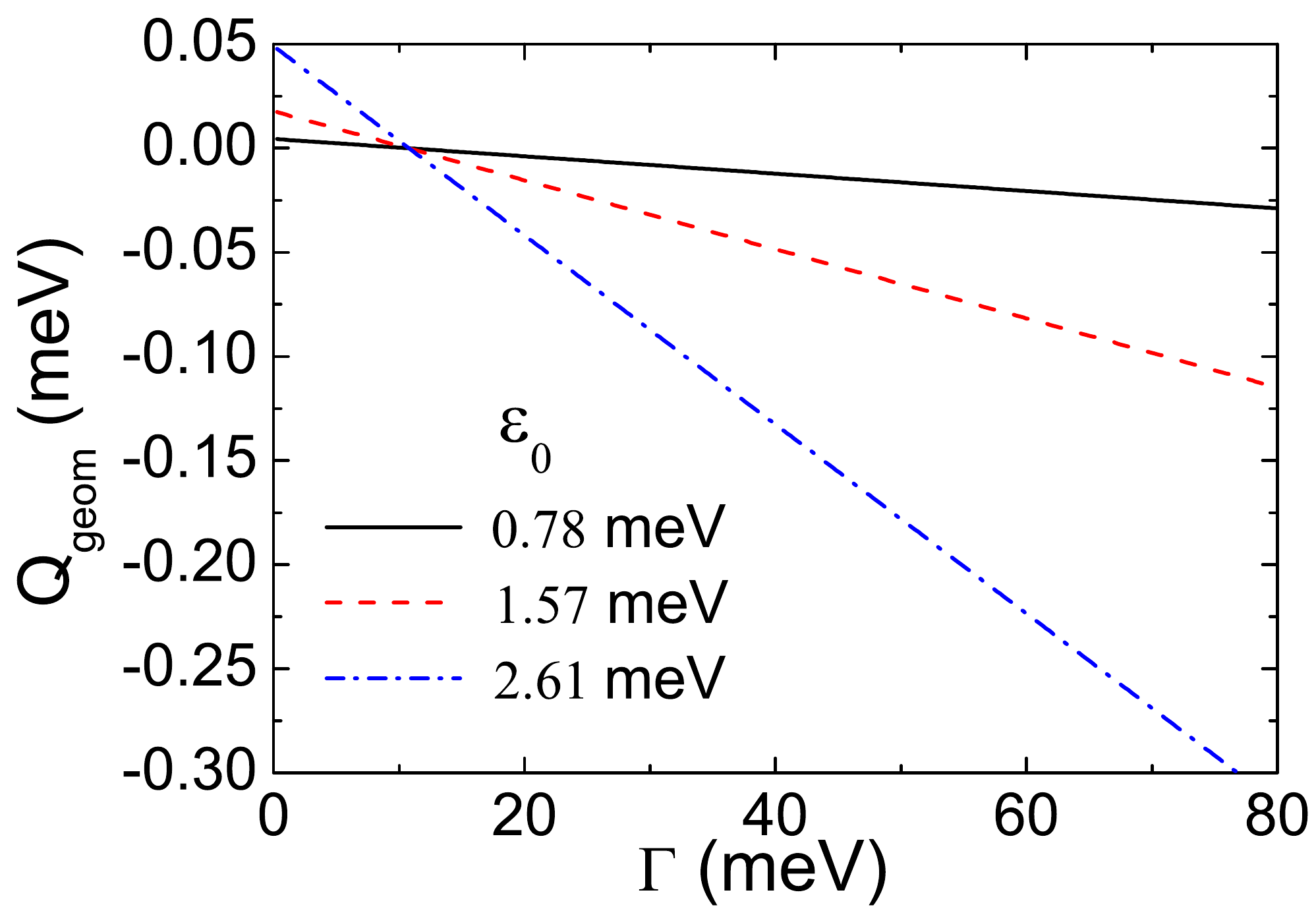}}
\vspace{-.4cm}
\caption{(Color online) Integrated geometric heat flux per period $Q_{geom}\equiv J_{geom}\mathcal T_p$ as a linear function of the system-bath coupling under the  two-bath-temperature modulation. We set the symmetric coupling $\Gamma_L=\Gamma_R=\Gamma$ and tunneling energy is $\Delta=5.22$ meV. When $\varepsilon_0\rightarrow-\varepsilon_0$, we observe the same lines.}
\label{fig:GTT}
\end{figure}

We first adiabatically modulate the two bath temperatures. The external control frequency $\Omega_{p}$ is chosen to be sufficiently small so that the adiabatic condition is respected \cite{Tp}. The protocol is chosen as $T_{L}(t)=150+90\cos(\Omega_{p}t)$ (K), $T_{R}(t)=150 +90\sin(\Omega_{p}t)$ (K), so that there is no temperature-bias-induced flux on average ($J_{dyn}=0$), but the geometric heat flux emerges. Figure \ref{fig:GTT} shows that the integrated geometric heat flux per period $Q_{geom}\equiv J_{geom}\mathcal T_p$ is linearly dependent on the system-bath coupling $\Gamma$. When $\Gamma$ approaches to zero, $Q_{geom}$ does not vanish but persists at some finite values. Also, we find that increasing the gap $\varepsilon_0$ can increase $Q_{geom}$ and when $\varepsilon_0\rightarrow-\varepsilon_0$, we observe the same lines. Generally, reversing the modulation cycle can reverse the direction of geometric heat flux. 
Moreover, as indicated in Fig. \ref{fig:GTT}, when the coupling strength $\Gamma$ exceeds some threshold values the geometric heat flux will also reverse its direction. This may offer a useful mean in dynamic control of the heat flux induced by adiabatic periodic modulation. 

\begin{figure}%[t]
\vspace{-.3cm}
\hspace{-.4cm}
\scalebox{0.42}[0.36]{\includegraphics{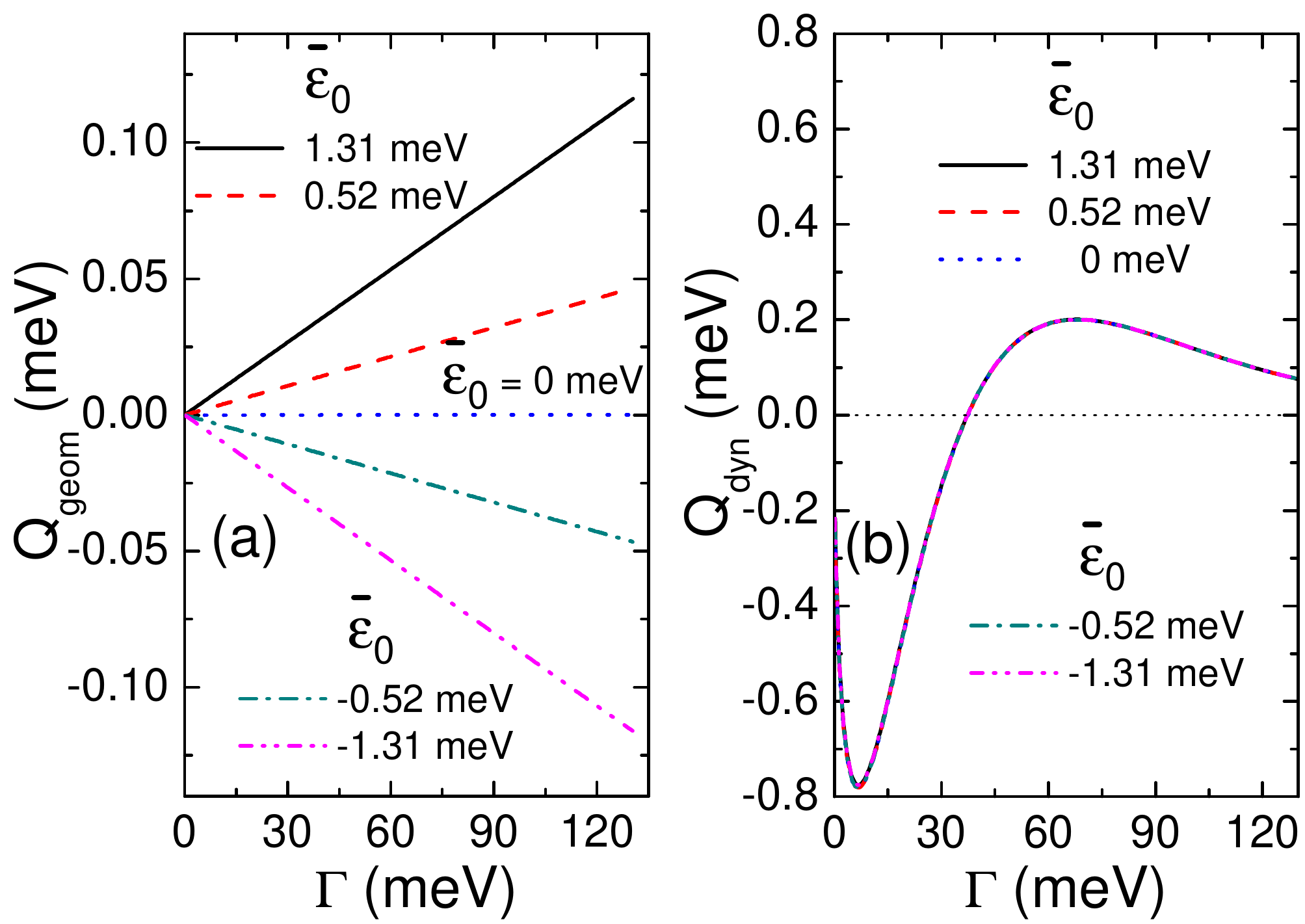}}
\vspace{-.4cm}
\caption{(Color online) (a) $Q_{geom}$ as a linear function of $\Gamma$ under modulations of one bath temperature and the gap $\varepsilon_0(t)$. We set $T_{L}(t)=150+90\cos(\Omega_p t)$ (K), $\varepsilon_{0}(t)=\overline{\varepsilon}_0+0.78\sin(\Omega_{p}t)$ (meV). $T_R=150$ K, $\Delta=5.22$ meV.
(b) Integrated dynamic heat flux per period $Q_{dyn}\equiv J_{dyn}\mathcal T_p$ under the same conditions for comparison.}
\label{fig:GTE}
\end{figure}

Second, we manipulate the bath temperature and the two-level energy gap $\varepsilon_{0}$.  As shown in Fig. \ref{fig:GTE}(a), $Q_{geom}$ is linearly dependent on the system-coupling strength. When $\Gamma$ approaches to zero, $Q_{geom}$ vanishes. Raising the average level-gap $\overline{\varepsilon}_0$ is able to increase the magnitude of $Q_{geom}$. If we reverse $\overline{\varepsilon}_0\rightarrow-\overline{\varepsilon}_0$, the geometric heat flux reverses its direction so that when $\overline{\varepsilon}_0=0$, $Q_{geom}$ is absent. Although in this modulation protocol the dynamic heat flux is nonzero, it decays as the system-bath coupling $\Gamma$ increases, [see Fig. \ref{fig:GTE}(b)]. In this way, the geometric heat flux will dominate the energy transport in the strong system-bath coupling regime.

\begin{figure}%[t]
%\hspace{-.2cm}
\scalebox{0.35}[0.35]{\includegraphics{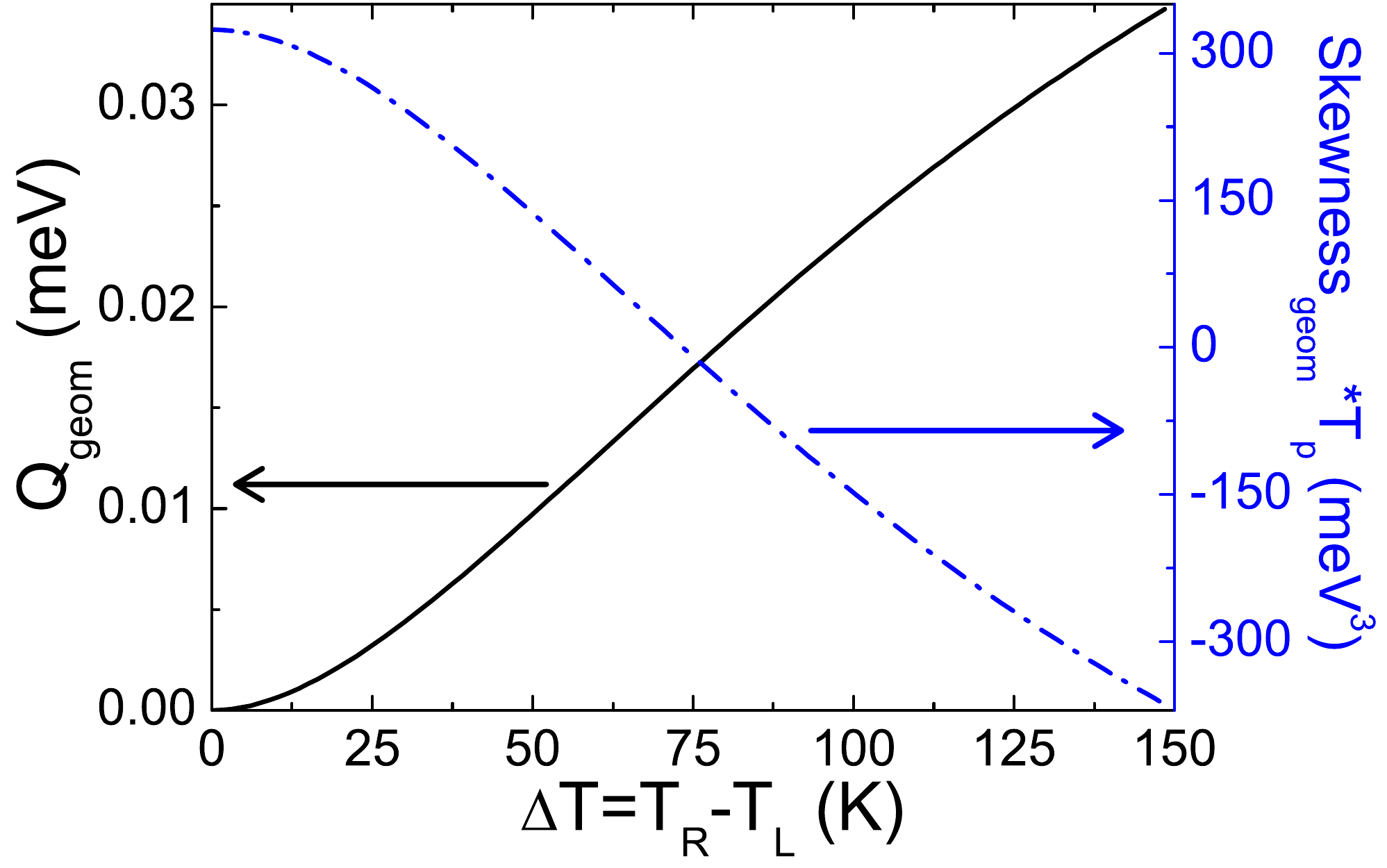}}
\vspace{-.4cm}
\caption{(Color online) Emergence of geometric phase effect and geometric heat flux for modulating two system-bath couplings. Although $Q_{geom}$ is absent at $T_L=T_R$, the nonzero geometric skewness ($\partial^3_{i\chi}\mathcal G_{geom}|_{\chi=0}$) shows the existence of the geometric phase effect, manifesting itself as the high-order heat transfer fluctuations. The control protocol is $\Gamma_{L}=130.5+104.4\cos(\Omega_p t)$ (meV) and $\Gamma_{R}=130.5+104.4\sin(\Omega_p t)$ (meV). Other parameters are $T_{L}=150$ K, $\omega_{0}=2.61$ meV and $\Delta=5.22$ meV. }
\label{fig:GEE}
\end{figure}

Besides these two control protocols discussed above, we finally consider the special case of modulating two system-bath couplings $\Gamma_L(t)$ and $\Gamma_R(t)$. Figure \ref{fig:GEE} shows the emergence of geometric heat flux when $T_L\neq T_R$. Although at $T_L=T_R$, the geometric heat flux vanishes, the geometric phase effect still exists, manifesting itself as the high-order heat transfer fluctuation, {\it e.g.}, the nonzero geometric skewness $\partial^3_{i\chi}\mathcal G_{geom}|_{\chi=0}\neq 0$. These observations are distinct from the previous results either in the quantum weak-coupling regime \cite{J. Ren} or in the classical regime \cite{S. Liu}, wherein under coupling-modulation, the Berry curvatures are always zero (no matter what the other parameter settings are), so that the geometric phase effect and geometric heat flux are always absent. It is the strong system-bath coupling  we consider here that makes the coupling-modulation-induced geometric heat pump nontrivial.

%{\it Impact on the heat flux fluctuation theorem.} 

In summary, we have studied the geometric phase-induced heat flux extensively in the spin-boson system under the adiabatic periodical modulation without the Redfield approximation. Using the NIBA approach, we have found that the geometric heat flux exists only when the two level system's energy gap is not zero. %This result shows that NIBA and the Redfield approximation can give different results for the same physical system even when the coupling is weak.  
Moreover, the pumping, no pumping, and dynamic control of the geometric heat flux have been discussed in detail. In particular, two-system-bath-coupling-modulation-induced geometric heat flux has been identified. So far, this nontrivial observation is exclusively for quantum transport in the strong system-bath coupling regime.

\section{acknowledgments}
{We thank Baowen Li for useful discussions. This work was supported in part by
the 10000-Plan of Shandong province, and the National High-Tech Program of China grant No. 2011AA010800 and 2011AA010803, NSFC grant No. 11174177 and 60725416. J.R. acknowledges the support of the U.S. Department of Energy through the LANL/LDRD Program.}

{ }

\begin{thebibliography}{ }
\bibitem{Tp} To satisfy the adiabatic condition, $\mathcal T_p=2\pi/\Omega_p$ should be much larger than the characteristic relaxation time of the whole system, including the dynamics of the two-level system and the baths. Also, $\mathcal T_p$ should be much larger than the relaxation time of the two population dynamics: $\mathcal T_c=1/(E_1-E_0)|_{\chi=0}$ \cite{S. Liu}. In our choice of parameters, the system's relaxation time $\mathcal T_c$ is on the order of picoseconds. Thus, the adiabatic condition can be satisfied when the modulation frequency $\Omega_p \ll 1$ THz. We choose $\mathcal T_p=1$ ns throughout the calculations.
\bibitem{braket} Different from the conventional Dirac's representation, the bra-ket notation we choose here just denotes the left-right eigenvector relation. They are not the conjugate transpose of each other.
\bibitem{note} The resonant  additive picture of the Redfield scheme is not the only possible physical picture in the weak-coupling limit. In fact, Ref.~\cite{L. A. Wu} pointed out that in the weak-coupling limit, by using a Born-Oppenheimer approach, one can obtain a different physical picture, with off-resonant tunneling and non-additive reservoir effects, which is similar to the NIBA scheme.
\end{thebibliography}
\end{document}